\documentclass[10pt,twocolumn,letterpaper]{article}

\usepackage{cvpr}
\usepackage{times}
\usepackage{epsfig}
\usepackage{graphicx}
\usepackage{amsmath}
\usepackage{amssymb}

\usepackage{algorithm,algpseudocode}
\usepackage{bm}
\usepackage{nicefrac}
\usepackage{multirow}
 
\usepackage[pagebackref=true,breaklinks=true,letterpaper=true,colorlinks,bookmarks=false]{hyperref}

 \cvprfinalcopy 

\def\cvprPaperID{9064} 

\ifcvprfinal\fi
\begin{document}

\title{A New Ensemble Method for Concessively Targeted Multi-model Attack}

\author{Ziwen He, Wei Wang, Xinsheng Xuan, Jing Dong, Tieniu Tan\\
Center for Research on Intelligent Perception and Computing,\\ National Laboratory of Pattern Recognition,\\
Institute of Automation, Chinese Academy of Sciences, Beijing, China\\
{\tt\small \{ziwen.he,xinsheng.xuan\}@cripac.ia.ac.cn,\{wwang,jdong,tnt\}@nlpr.ia.ac.cn}
}

\maketitle

\begin{abstract}
    It is well known that deep learning models are vulnerable to adversarial examples crafted by maliciously adding perturbations to original inputs. There are two types of attacks: targeted attack and non-targeted attack, and most researchers often pay more attention to the targeted adversarial examples. However, targeted attack has a low success rate, especially when aiming at a robust model or under a black-box attack protocol. In this case, non-targeted attack is the last chance to disable AI systems. Thus, in this paper, we propose a new attack mechanism which performs the non-targeted attack when the targeted attack fails. Besides, we aim to generate a single adversarial sample for different deployed models of the same task, e.g. image classification models. Hence, for this practical application, we focus on attacking ensemble models by dividing them into two groups: easy-to-attack and robust models. We alternately attack these two groups of models in the non-targeted or targeted manner. We name it a bagging and stacking ensemble (BAST) attack. The BAST attack can generate an adversarial sample that fails multiple models simultaneously. Some of the models classify the adversarial sample as a target label, and other models which are not attacked successfully may give wrong labels at least. The experimental results show that the proposed BAST attack outperforms the state-of-the-art attack methods on the new defined criterion that considers both targeted and non-targeted attack performance. 
    
\end{abstract}

\begin{figure}[!h!t]
	\begin{center}
		\includegraphics[width=1.0\linewidth]{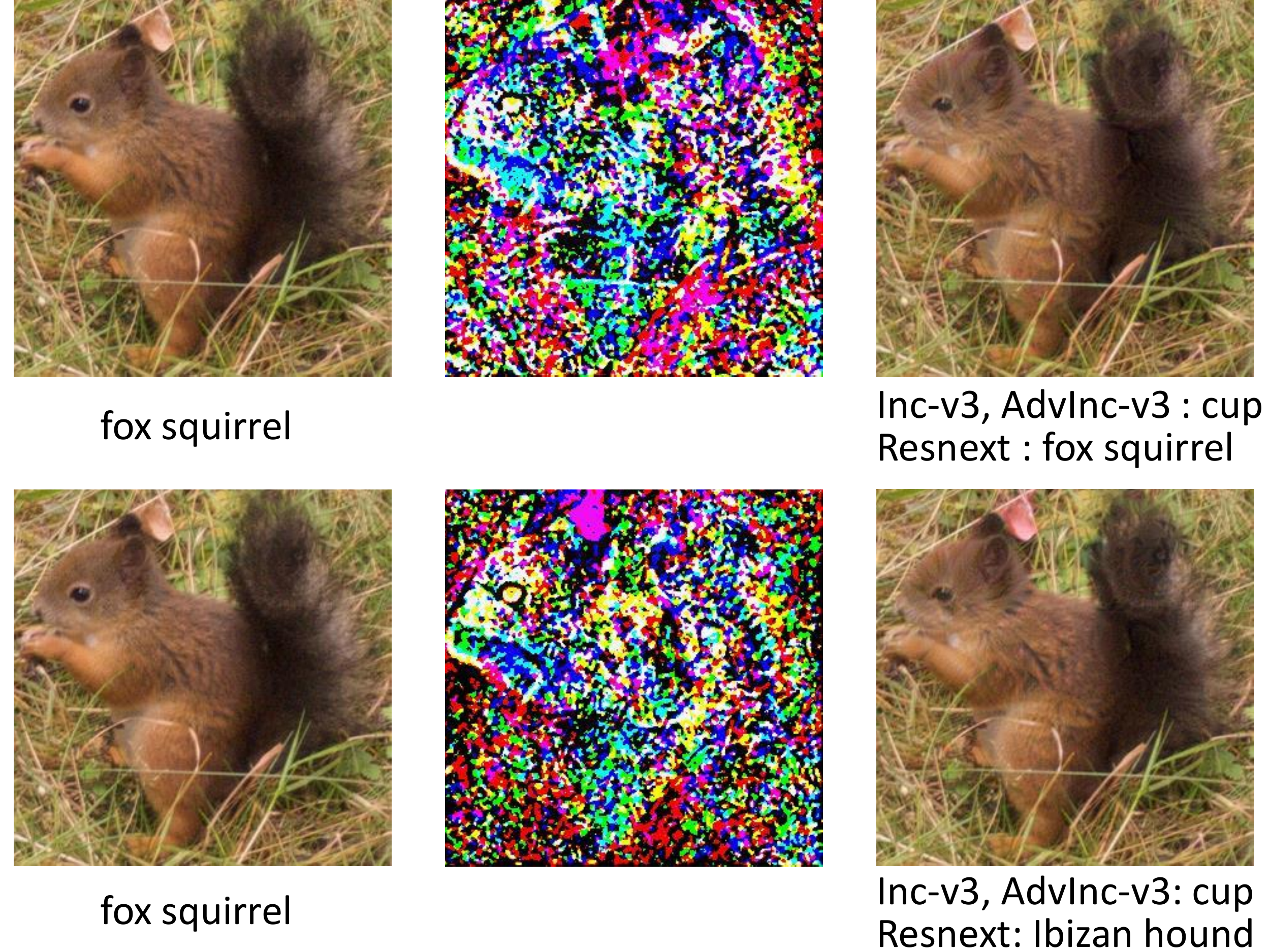}
	\end{center}
	\caption{Adversarial examples generated for
		Inc-v3~\cite{Szegedy2015RethinkingTI}, AdvInc-v3~\cite{Tramr2017EnsembleAT} and ResnextDenoiseAll~\cite{Xie2018FeatureDF} under the white-box protocol. \textbf{Left column}: the original images. \textbf{Middle column}: the adversarial noises. \textbf{Right column}: adversarial images. \textbf{Top row}: ensemble in loss method~\cite{Dong2017BoostingAA}. \textbf{Bottom row}: our method. Both adversarial images are classified by Inc-v3 and AdvInc-v3 as the target label ``cup''. While the adversarial image in top row can't fool ResnextDenoiseAll, the adversarial image in bottom row generated by our method successfully fools ResnextDenoiseAll, classified by ResnextDenoiseAll as ``Ibizan hound''.}
	\label{fig:problem}
\end{figure}
\section{Introduction}

Recent research has shown that machine learning models can easily be fooled by adversarial samples which are crafted by adding designed perturbations to the inputs~\cite{Goodfellow2014ExplainingAH,Szegedy2013IntriguingPO}. From the perspective of attack goals, there are two types of adversarial attacks: (1) the non-targeted attack, which crafts adversarial examples misclassified as wrong labels; (2) the targeted attack, which generates adversarial examples classified as target labels. People are more interested in the targeted attack. However, the targeted attack has a lower success rate.

In a black-box attack protocol, it is probable to utilize the transferability of crafted adversarial examples which can be misclassified by unseen target models~\cite{Papernot2016TransferabilityIM,Liu2016DelvingIT,Papernot2016PracticalBA}. While existing approaches are effective to generate non-targeted transferable adversarial examples, targeted adversarial examples are hardly transferred~\cite{Liu2016DelvingIT}. Even in a white-box attack protocol, the targeted attack also has little effect on some adversarial robust models. As shown in Fig.\ref{fig:problem}, targeted attacks on three models are conducted in the top row, with an ensemble attack method in \cite{Dong2017BoostingAA}. Inc-v3~\cite{Szegedy2015RethinkingTI} and AdvInc-v3~\cite{Tramr2017EnsembleAT} miss-classify the adversarial image as the target label ``cup'', while the adversarial robust model ResnextDenoiseAll~\cite{Xie2018FeatureDF} still successfully classify it as the true label, which means the adversarial robust model is hard to be targetedly attacked, even for the white-box attack. 

In this paper, we propose a new adversarial attack mechanism, which maximizes the attack success rate by considering both the targeted and non-targeted attacks but prioritizing targeted attack. 
When the targeted attack fails, this adversarial sample hopefully performs the non-targeted attack successfully. Besides, as shown in the bottom row of Fig.\ref{fig:problem}, the generated single adversarial example is able to fool all deployed models and even for ResNextDenoiseAll, the non-targeted attack at least succeed. 
Thus in our attack, different models are firstly divided into two groups, easy-to-attack models and robust models. Two groups of models are then alternately attacked. We name this attack a bagging and stacking ensemble (BAST) attack. Our contributions are as follows:

1. We present a new adversarial attack mechanism, at-least-non-targeted attack. To test the efficiency of methods in this attack protocol, we design a new evaluation criterion.

2. We propose a novel ensemble attack method, BAST attack, to make sure at-least-non-targeted attack succeed. BAST attack outperforms the state-of-the-art attack methods with respect to at-least-non-targeted attack.

3. Experimental results on image classification models demonstrate that adversarial examples crafted by our proposed method are able to fool multiple models in either white-box or black-box protocol, achieving an optimal trade-off between non-targeted attack and targeted attack.

\section{Backgrounds}

In this section, we review the backgrounds of adversarial attack and defense methods. Following Sharma \etal~\cite{Sharma2018CAAD2G}, we conduct attack based on MIFGSM~\cite{Dong2017BoostingAA}, utilizing some methods in sec.~\ref{sec:attack methods}, including input diversity~\cite{Xie2018ImprovingTO} and a translation-invariant method~\cite{Dong2019EvadingDT}, to improve the transferability. The defense models to evaluate our proposed method include pretrained models on ImageNet~\cite{Russakovsky2014ImageNetLS} and some robust models trained with some defense methods in sec.~\ref{sec:defense_methods}.

\subsection{Attack Methods}
\label{sec:attack methods}

\subsubsection{Fast Gradient Sign Method}

Fast gradient sign method (FGSM)~\cite{Szegedy2013IntriguingPO} performs a single step update on the original sample $\bm{x}$ along the direction of the gradient of a loss function $J(\bm{x};y;\theta)$, where $J$ is often the cross-entropy loss. In the scenario where perturbation $\epsilon$ is to meet the $L_{\inf}$ norm bound $\parallel \bm{x}^{*} - \bm{x} \parallel_{\infty} < \epsilon $, the adversarial example is computed as 
\begin{equation}
\bm{x}^* = \bm{x} + \epsilon\cdot\mathrm{sign}(\nabla_{\bm{x}}J(\bm{x},y))
\label{eq:fgsm}
\end{equation}

\subsubsection{Momentum-based Iterative Method}

To solve the problem that most adversarial attacks fool black-box models with low success rate, Dong \etal propose a momentum-based iterative algorithm, MI-FGSM~\cite{Dong2017BoostingAA}. By integrating momentum term into the iterative process of attack, this method can craft more transferable adversarial examples by computing as
\begin{equation}
\label{eq:mim1}
\bm{g}_{t+1} = \mu \cdot \bm{g}_{t} + \frac{\nabla_{\bm{x}}J(\bm{x}_{t}^*,y)}{\|\nabla_{\bm{x}}J(\bm{x}_{t}^*,y)\|_1}
\end{equation}
\begin{equation}
\label{eq:mim2}
\bm{x}_{t+1}^* = \bm{x}_{t}^* + \alpha\cdot\mathrm{sign}(\bm{g}_{t+1})
\end{equation}

\subsubsection{Input Diversity Method}

Xie \etal propose that diverse inputs can improve transferability of adversarial examples~\cite{Xie2018ImprovingTO}. Based on the MI-FGSM~\cite{Dong2017BoostingAA}, random transformation is performed on the input during each iteration to realize the input diversity. Their experiments show that the combination of random scaling and random zero padding has the best performance.

\subsubsection{Translation-invariant Method}

Dong \etal propose a translation-invariant attack method to generate more transferable adversarial examples against defense models~\cite{Dong2019EvadingDT}. To improve the efficiency, the translation operation is realized by convolving the gradient of the untranslated image with a predefined kernel function
$W$. The update rule to compute adversarial examples is as following:
\begin{equation}
\label{eq:tim}
\bm{x}^* = \bm{x} + \alpha\cdot\mathrm{sign}(\bm{W}\ast\nabla_{\bm{x}}J(\bm{x},y))
\end{equation}

\subsubsection{Ensemble Method}

 Ensemble methods have been widely adopted in previous researches to enhance the performance of neural networks~\cite{Hansen1990NeuralNE, Caruana2004EnsembleSF, Krogh1994NeuralNE}. For example, Bagging~\cite{Breiman1996BaggingP} and Stacking~\cite{WolpertStacked}, can both improve accuracy and robustness of neural networks. Recently, Liu \etal propose novel ensemble-based approaches to generate adversarial examples, which improve the transferability even for targeted adversarial examples~\cite{Liu2016DelvingIT}. 

\subsection{Defense Methods}
\label{sec:defense_methods}

Adversarial training~\cite{Szegedy2013IntriguingPO,Goodfellow2014ExplainingAH} is the simplest and most widely used method to defense adversarial attack. It can increase robustness by directly adding a considerable amount of adversarial examples generated by different attack methods to the training set during network training. Madry \etal~\cite{Madry2017TowardsDL} regard adversarial training as a framework of maximum and minimum game and train more robust models in this way. 

Xie \etal propose a denoising network architecture~\cite{Xie2018FeatureDF}, which enhances adversarial robustness by adding feature denoising module. 
Combined with adversarial training, feature denoising networks greatly improve the adversarial robustness, especially in targeted attack protocol. However, non-targeted attack is still a threat to these defense methods.


\begin{algorithm}[t]
	\small
	\caption{Attack on single model}
	\label{alg:attack}
	\begin{algorithmic}[1]
		\Require A classifier $f$ with loss function $J$; a real example $\bm{x}$ and ground-truth label $y$; The size of perturbation $\epsilon$; iterations $T$ and decay factor $\mu$; The random preprocess function $div()$; gaussian kernel W. 
		\Ensure
		An adversarial example $\bm{x}^*$ with $\|\bm{x}^* - \bm{x}\|_{\infty} \leq \epsilon$.
		\State $\alpha = \nicefrac{\epsilon}{T}$; $\bm{g}_0 = 0$; $\bm{x}_0^* = \bm{x}$;
		\For {$t = 0$ to $T-1$}
		\State Do random preprocess for input to obtain $div(\bm{x}_t^*)$;
		\State Input processed image $div(\bm{x}_t^*)$ to $f$ and obtain the gradient $\nabla_{\bm{x}}J(div(\bm{x}_t^*),y)$;
		\State Convolve the gradient with $\bm{W}$ to get smoothed gradient $\bm{W}\ast \nabla_{\bm{x}}J(div(\bm{x}_t^*),y)$;
		\State Update $\bm{g}_{t+1}$ by accumulating the velocity vector in the gradient direction as 
		\vspace{-2ex}
		\begin{equation}
		\label{eq:momentum}
		\bm{g}_{t+1} = \mu \cdot \bm{g}_{t} + \frac{\bm{W}\ast\nabla_{\bm{x}}J(div(\bm{x}_t^*),y)}{std(\bm{W}\ast\nabla_{\bm{x}}J(div(\bm{x}_t^*),y))};
		\end{equation}
		\vspace{-3ex}
		\State Update $\bm{x}_{t+1}^*$ by applying the clipped gradient as
		\vspace{-1.5ex}
		\begin{equation}
		\label{eq:update}
		\bm{x}_{t+1}^* = \bm{x}_{t}^* + \alpha\cdot\mathrm{clip}_{[-2,2]}(round(\bm{g}_{t+1}));
		\end{equation}
		\vspace{-4ex}
		\EndFor \\
		\Return $\bm{x}^* = \bm{x}_T^*$.
	\end{algorithmic}
\end{algorithm}

\section{The Proposed Method}

To generate adversarial examples in at-least-non-targeted protocol, we propose a novel ensemble attack method, BAST attack. In this section, we first give a brief introduction of previous ensemble methods and explain their drawbacks in at-least-non-targeted attack protocol. We then introduce the attack method for a single model, which is the base for our ensemble attack. Finally, we present our solution, BAST attack, which enables us to efficiently craft adversarial examples in at-least-non-targeted attack protocol.

\subsection{Previous Ensemble Methods}
\label{sec:pa-ensemble}

In NIPS 2017 adversarial attack competition~\cite{Kurakin2018AdversarialAA}, Dong \etal~\cite{Dong2017BoostingAA} report three different ensemble methods including \textit{ensemble in logits}, \textit{ensemble in predictions} and \textit{ensemble in loss}, the only difference of which is where to combine the outputs of multiple models. All methods simply add model's outputs together and then average.
We focus on ensemble in loss and give its formulation as follows:

 \begin{equation}\label{eq:loss}
 J(\bm{x}, y) = \textstyle\sum_{i=1}^N w_i J_i(\bm{x}, y)
 \end{equation}
 where $J_i(\bm{x}, y)$ is the cross-entropy loss of the $\textit{i}$-th model. 
 
\begin{figure}
	\begin{center}
		\includegraphics[width=1.0\linewidth]{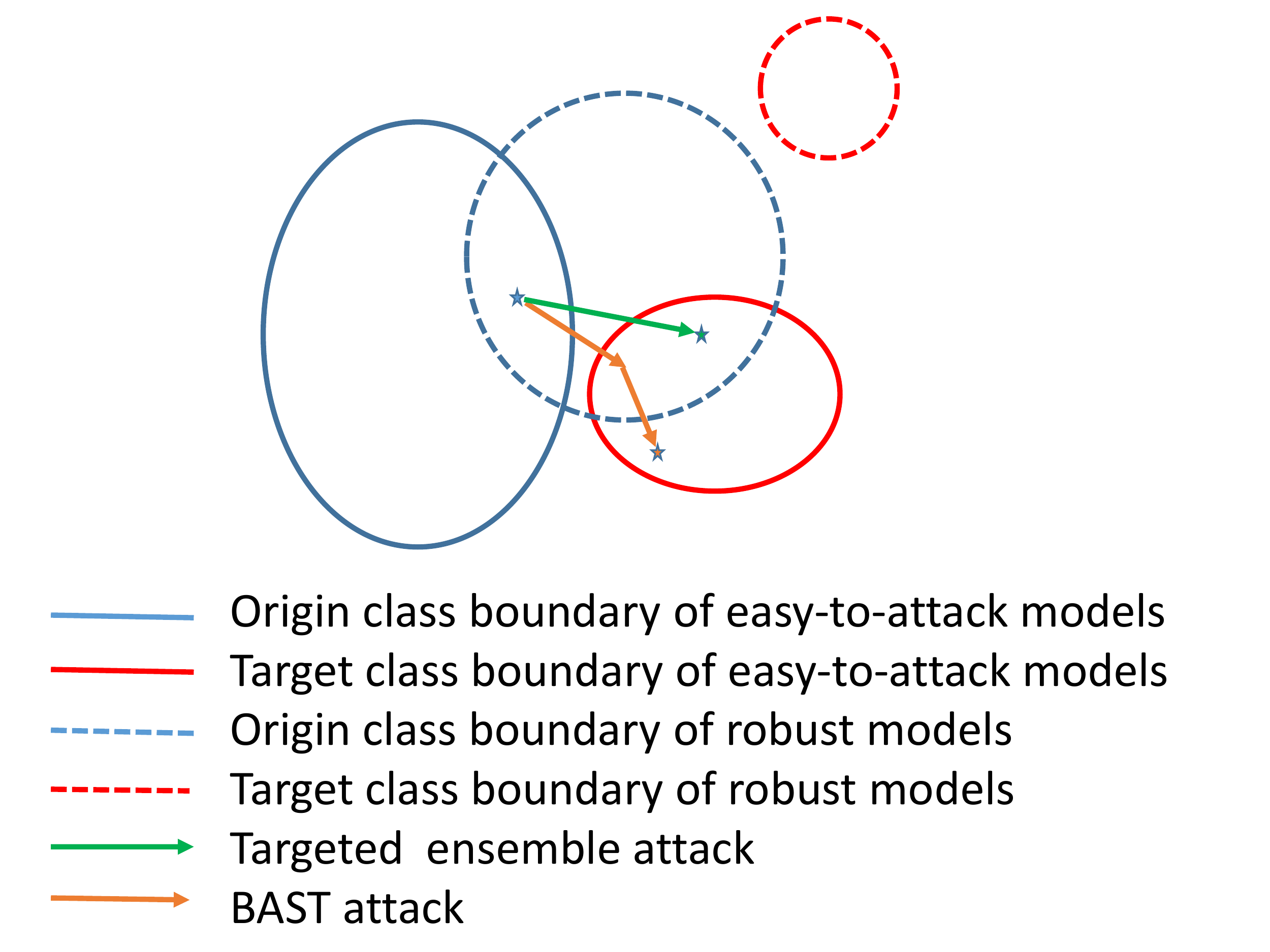}
	\end{center}
	\caption{Illustration of targeted ensemble attack versus BAST attack.}
	\label{fig:explain}
\end{figure}

 The loss function $J(\bm{x}, y)$ is optimized by gradient-based algorithms such as FGSM. When the targeted attack is performed on the ensemble, the gradient always points to target class boundary of easy-to-attack models rather than that of robust models, shown as the green arrow in Fig.~\ref{fig:explain}. To fully utilize the the gradient information of robust models, we propose our method, BAST attack.
  
 

\begin{figure}[t]
	\begin{center}
		\includegraphics[width=1.0\linewidth]{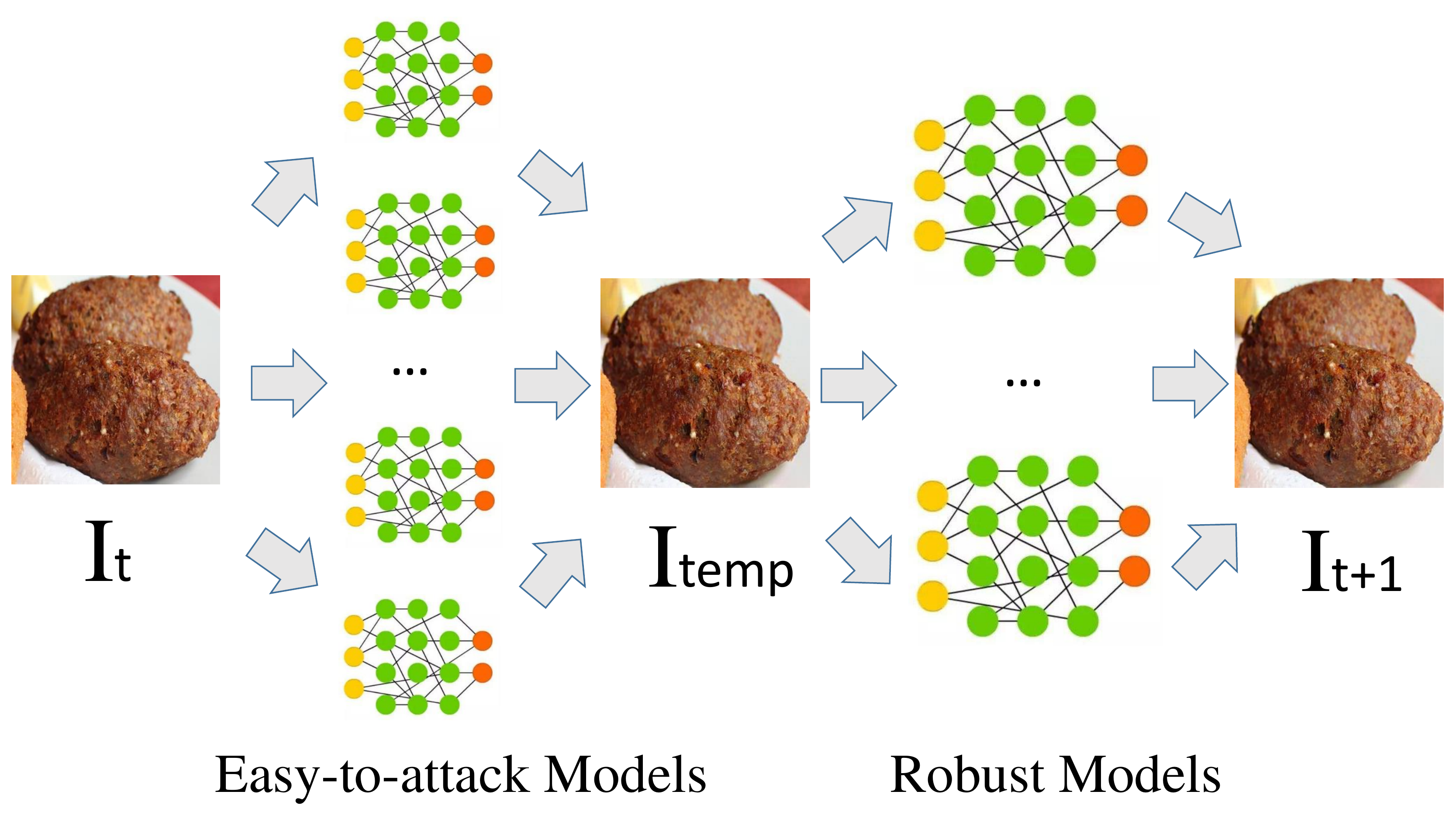}
	\end{center}
	\caption{One iteration of \textbf{BAST attack}. It has two main steps: (1) From $I_t$ to $I_{temp}$, attacks are performed on easy-to-attack models. (2)From $I_{temp}$ to $I_{t+1}$, attacks are performed on robust models.}
	\label{fig:long}
\end{figure}

\subsection{Attack on Single Model}

To achieve high success rate and strong transferability of adversarial examples, we follow Sharma \etal~\cite{Sharma2018CAAD2G} to use Algorithm~\ref{alg:attack} as our baseline attack for a single model. For targeted attack, ground-truth label $y$ is exchanged with target class label $y^{*}$ and the plus sign in Eq.(\ref{eq:update}) is changed to the minus sign. To further improve the transferability in black-box attack scenario, cropping is added into the random preprocess~\cite{Xie2018ImprovingTO}.  

\subsection{BAST Attack}
 
 Motivated by ensemble methods such as Bagging~\cite{Breiman1996BaggingP} and Stacking~\cite{WolpertStacked}, we propose a novel ensemble method for adversarial attack, called bagging and stacking ensemble (BAST) attack, as shown in Algorithm ~\ref{alg:pase-ensemble-attack}.
 
 In our BAST attack, all models are divided into two groups: easy to attack and robust models. The models in each group compose an ensemble model in a way similar to Bagging. The same type of attack, non-targeted or targeted attack, is performed on each ensemble. The Bagging ensemble model obtains a lower variance by averaging predictions of independent models. In another aspect, different groups compose an ensemble model in a way similar to Stacking. In Stacking, outputs returned by some lower layer weak learners are used to train a meta model, while in our BAST attack, outputs of last Bagging ensemble model are fed into the next one. 
 
 Fig.~\ref{fig:long} shows one iteration of BAST attack. 
 In practical application, we conduct targeted attack on easy-to-attack models for $m$ times and conduct non-targeted attack on other models for $n$ times, respectively. By controlling these two hyperparameters, the performance of BAST attack can be further improved. 
 

\begin{algorithm}[t]
	\small
	\caption{BAST attack}
	\label{alg:pase-ensemble-attack}
	\begin{algorithmic}[1]
		\Require The logits of classifiers $\bm{l}_1, \bm{l}_2, ..., \bm{l}_{N+M}$; ensemble weights $w_1, w_2, ..., w_{N+M}$; non-targeted attack times $n$ and targeted attack times $m$; a real example $\bm{x}$ and ground-truth label $y_{ture}$, target label $y_{target}$; the size of perturbation $\epsilon$; iterations $T$ and decay factor $\mu$; the random preprocess function $div()$; gaussian kernel W.
		\Ensure
		An adversarial example $\bm{x}^*$ with $\|\bm{x}^* - \bm{x}\|_{\infty} \leq \epsilon$.
		\State $\alpha = \nicefrac{\epsilon}{T}$; $\bm{g}_0 = 0$; $\bm{x}_0^* = \bm{x}$;
		\For {$t = 0$ to $T-1$}
		\State $\bm{x}_{t+1}^*=\bm{x}_{t}^*$
		\For {$i = 0$ to $1$}
		\If {i=0}
		\State $k_0=1$, $K=N$, $y=y_{true}$, $s=n$;
		\Else
		\State $k_0=N+1$, $K=N+M$, $y=y_{target}$, $s=m$;
		\EndIf
		\For {$j = 0$ to $s-1$}
		\State $\bm{x}_{t}^*=div(\bm{x}_{t+1}^*)$
		\State Input $\bm{x}_t^*$ and output $\bm{l}_k(\bm{x}_t^*)$ for $k = k_0, k_0+1, ..., K$;
		\State  Get softmax cross-entropy loss $J_k(\bm{x}_t^*,y)$ based on $J_k(\bm{x}_t^*, y) = -\mathbf{1}_{y}\cdot\log(\mathrm{softmax}(\bm{l_k}(\bm{x}_t^*)))$;
		\State Total loss $J(\bm{x}_t^*, y) = \sum_{k=k_0}^Kw_kJ_k(\bm{x}_t^*, y)$;
		\State Obtain the gradient $\nabla_{\bm{x}}J(\bm{x}_t^*,y)$;
		\State Update $\bm{g}_{t+1}$ by Eq.~\eqref{eq:momentum};
		\State Update $\bm{x}_{t+1}^*$ by Eq.~\eqref{eq:update};
		\EndFor
		\EndFor 
		\EndFor \\
		\Return $\bm{x}^* = \bm{x}_T^*$.
	\end{algorithmic}
\end{algorithm}

\section{Experiments}

In this section, we conduct extensive experiments to validate the effectiveness of
the proposed method. We first specify the experimental settings in Sec.~\ref{sec:setup}. Then we describe a new evaluation criterion in at-least-non-targeted attack protocol in Sec.~\ref{sec:evaluation}. We further evaluate our proposed method utilizing different adversarial models in Sec.~\ref{sec:main-results} and Sec.~\ref{sec:strategy}. 

\subsection{Setup}
\label{sec:setup}

We conduct experiments on ImageNet~\cite{Russakovsky2014ImageNetLS}. The maximum perturbation $\epsilon$ is set to 16, with pixel values in [0, 255].
We ensemble three models, which are a normally trained model---Inception v3 (Inc-v3)~\cite{Szegedy2015RethinkingTI}, an adversarially trained model---AdvInception v3 (AdvInc-v3)~\cite{Tramr2017EnsembleAT} and an extremely robust model trained by Facebook---ResnextDenoiseAll (AdvDeRex)~\cite{Xie2018FeatureDF}, as the substitute model to be attacked. For BAST attack, we set Inc-v3 and AdvInc-v3 as easy-to-attack models, AdvDeRex as the robust model. Besides white-box attack success rates of the three models, we also report black-box attack success rates evaluated on a normally trained model---Inception v1 (Inc-v1)~\cite{Szegedy2014GoingDW} and an adversarially trained model---AdvInceptionResnet v2 (AdvIncres-v2)~\cite{Tramr2017EnsembleAT}. We evaluate with 1000 images from ImageNet~\cite{Russakovsky2014ImageNetLS} which are specially chosen to be classified by our models correctly. For traditional ensemble attack methods, the number of iterations is 100, which is enough to make the algorithm converge. 
For BAST attack, as we conduct targeted attack on easy-to-attack models for $m$ times and then conduct non-targeted attack on robust models for $n$ times, the number of iterations is set as [$200/(m+n)$] for comparison. We set $m=2$ and $n=1$.

			


\begin{table*}[!ht]
	\begin{center}
		\begin{tabular}{|l|c|c|c|c|c|}
			\hline
			Method & Inc-v3 & AdvInc-v3 & AdvDeRex & Inc-v1 * & AdvIncres-v2 * \\
			\hline\hline

			Bagging attack & 0.0/100.0/100.0 & 0.0/99.8/\textbf{99.8} & 18.3/4.0/13.15 & 32.7/6.2/22.55 & 23.7/7.5/19.35\\
			\hline
			Stacking attack & 0.0/99.5/99.5 & 3.0/91.2/92.7 & 4.9/0.1/2.55 & 11.3/0.9/6.55 & 14.0/0.5/7.5\\
			\hline
			
			BAST attack & 0.0/100.0/\textbf{100.0}& 0.7/98.2/98.55 & 72.0/0.1/\textbf{36.1} & 40.5/6.7/\textbf{26.95} & 29.7/8.5/\textbf{23.35} \\
			\hline
		\end{tabular}
	\end{center}
	\caption{Results on ImageNet, shown as A/B/C(\%), where A represents that non-targeted attack succeeds but targeted attack fails, B represents targeted attack succeeds and C is the score computed by Eq.(\ref{eq:evaluation}). The sign * represents attack on this model is black-box attack.}
	\label{tab:at-least-non-targeted}
\end{table*}

\begin{figure*}[!h!t]
	\begin{center}
		\includegraphics[width=1.0\linewidth]{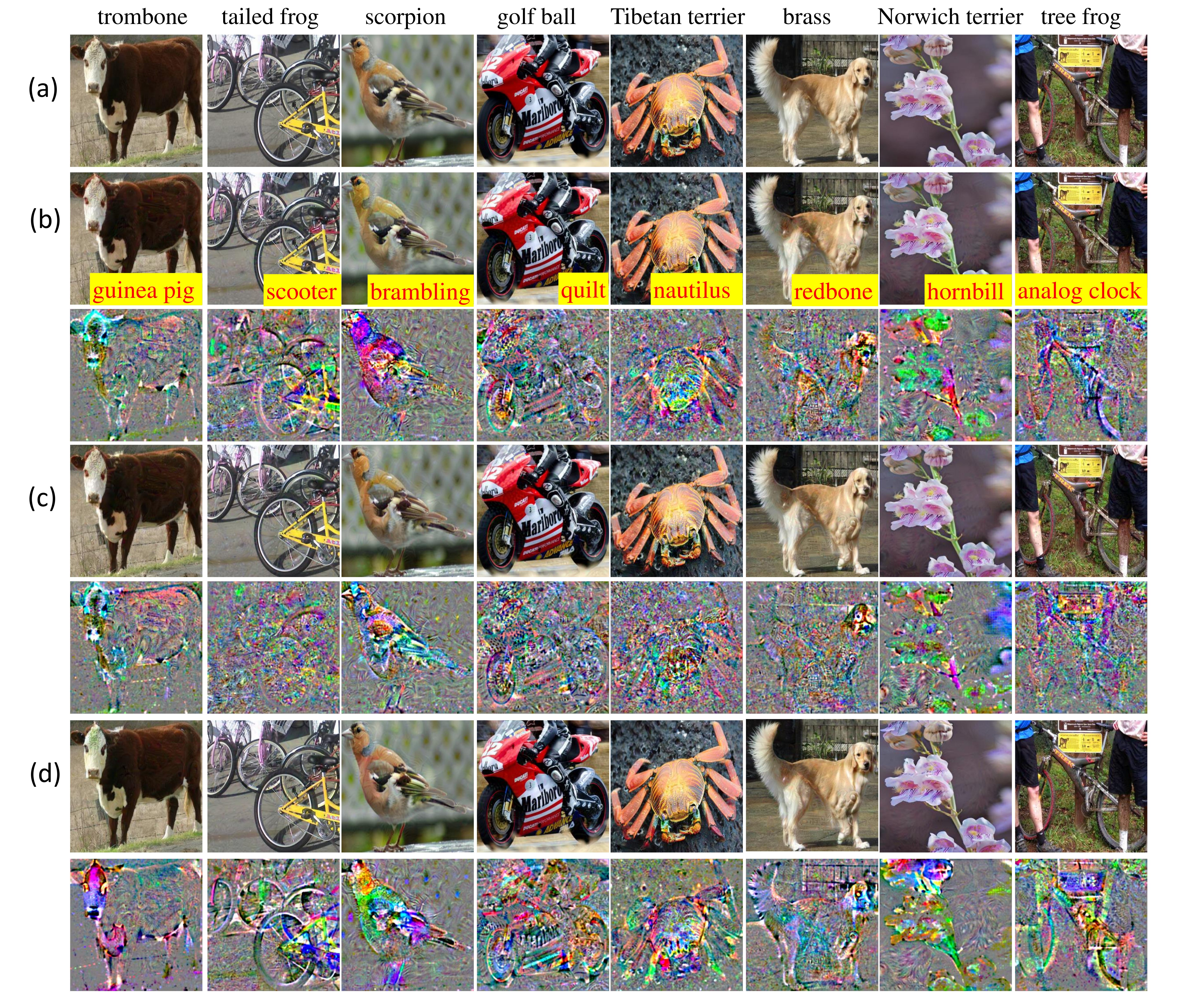}
	\end{center}
	\caption{Adversarial examples and perturbations generated by BAST attack, Bagging attack and Stacking attack. \textbf{(a)}: Natural examples and their target class labels. \textbf{(b)}: BAST attack. The predicted labels by AdvDeRex are in the yellow patch. \textbf{(c)}: Bagging attack. \textbf{(d)}: Stacking attack. All adversarial images are classified by Inc-v3 and AdvInc-v3 as the target labels, while only adversarial images in (b) successfully fool AdvDeRex.}
	\label{fig:examples}
\end{figure*}

\subsection{Evaluation Criteria}
\label{sec:evaluation}

Notice that we aim to obtain a weighted combination of targeted and non-targeted attack. Thus a new evaluation criterion is needed to verify the efficiency of tested methods. The evaluation score is defined as the following formula: 

\begin{equation}
\label{eq:evaluation}
F = \frac{1}{n}\sum_{i=1}^n F(x_i)
\end{equation}
and 
\begin{equation}
\label{eq:F}
F(x_i) =\left\{
             \begin{array}{lr}
          1  ,  \: \: \: \: \: if \, targeted \, attack \, succeeds &  \\

          0.5 , \: \: if \, only \, nontargeted\, attack \,succeeds & \\
          0 , \: \: \: \: \: if \, attack \, fails
             \end{array}
\right.
\end{equation}
where n is the number of images for evaluation and $F$ is the attack score on one model.
For each image $x_i$, if targeted attack is successful on a model, $F(x_i)$ is 1 point on this model. If targeted attack is not successful, but non-targeted attack is successful, then $F(x_i)$ is 0.5 point. Otherwise, $F(x_i)$ is 0 point.

%

\begin{table*}
	\begin{center}
		\begin{tabular}{|l|c|c|c|c|c|c|}
			\hline
			Method & Inc-v3 & AdvInc-v3 & AdvDeRex & Inc-v1 * & AdvIncres-v2 * \\
			\hline\hline
			
			Without-stacking & 0.0/100.0/\textbf{100.0} & 0.0/99.9/\textbf{99.9} & 62.5/0.1/31.35 & 41.3/3.7/24.35 & 32.0/5.0/21.0 \\
			\hline
			Without-bagging & 0.0/99.1/99.1 & 2.8/93.5/94.9 & 69.7/0.1/34.95 & 39.3/5.0/24.65 & 27.7/5.3/19.15 \\
			\hline
			BAST & 0.0/100.0/\textbf{100.0} & 0.7/98.2/98.55 & 72.0/0.1/\textbf{36.1} & 40.5/6.7/\textbf{26.95} & 29.7/8.5/\textbf{23.35}\\
			\hline
			
		\end{tabular}
	\end{center}
	\caption{Results for different ensemble methods, shown as A/B/C. A represents that non-targeted attack succeeds but targeted attack fails, B represents targeted attack succeeds and C is the score computed by Eq.(\ref{eq:evaluation}). The sign * represents attack on this model is black-box attack.}
	\label{tab:compare}
\end{table*}
\begin{table*}
	\begin{center}
		\begin{tabular}{|l|c|c|c|c|c|}
			\hline
			Hyperparameters & Inc-v3 & AdvInc-v3 & AdvDeRex & Inc-v1 * & AdvIncres-v2 * \\
			\hline\hline
			(a) m=1, n=1  & 0.0/99.6/99.6 & 2.8/92.2/93.6 & 77.7/0.1/38.95 & 41.6/3.7/24.5 & 30.8/5.4/20.8 \\
			(b) m=2, n=1  & 0.0/100.0/100.0 & 0.7/98.2/98.55 & 72.0/0.1/36.1 & 40.5/6.7/26.95 & 29.7/8.5/23.35\\
			(c) m=3, n=1 & 0.0/100.0/100.0 & 0.3/98.9/99.05 & 67.3/0.1/33.75 & 39.8/7.5/27.4 & 27.1/9.9/23.45\\
			(d) m=5, n=5 & 0.0/99.8/99.8 & 3.0/92.0/93.5 & 74.4/0.1/37.3 & 38.8/4.0/23.4 & 28.6/4.2/18.5 \\
			(e) m=10, n=10 & 0.0/99.9/99.9 & 3.4/91.6/93.3 & 71.6/0.1/35.9 & 37.5/4.2/22.95 & 27.0/4.6/18.1\\
			\hline
		\end{tabular}
	\end{center}
	\caption{Results for different combinations of $m$ and $n$ in BAST attack, shown as A/B/C. A represents that non-targeted attack succeeds but targeted attack fails, B represents targeted attack succeeds and C is the score computed by Eq.(\ref{eq:evaluation}). The sign * represents attack on this model is black-box attack.}
	\label{tab:strategy}
\end{table*}

\subsection{Main Results}
\label{sec:main-results}

We evaluate three methods: (1) Bagging attack. The targeted attack is performed on all three models in the bagging way, which means the ensemble method is Eq.(\ref{eq:loss}), where the ensemble weight $w_i$ for each model is the same. (2) Stacking attack. The targeted attack is performed on each model alternately in the stacking way. (3) Our BAST attack. 

Our main results are in Table~\ref{tab:at-least-non-targeted}. We show that:

(1) Our new attack mechanism, at-least-non-targeted attack, plays a key role in attacking multiple models, especially when attacking some robust models. 

Focus on the column ``AdvDeRex'' in Table~\ref{tab:at-least-non-targeted}. When we conduct targeted ensemble attack, the highest targeted attack success rate on AdvDeRex is 4.0\%. The low success rates indicate that targeted adversarial examples have a low attack success rate especially on robust models. Thus a successful non-targeted attack on these robust models is essential. 

(2) Our BAST attack outperforms other methods in the at-least-non-targeted attack protocol. 

BAST attack gets competitive results with other two targeted ensemble attacks on easy-to-attack models. On the robust model, however, BAST attack largely improve the non-targeted attack success rate with little decrease in targeted attack success rate. We list some natural images in ImageNet and corresponding adversarial images and perturbations crafted by BAST attack in Fig.~\ref{fig:examples}.  Moreover, in black-box attack on ImageNet, BAST attack outperforms targeted ensemble attacks on the success rate of both non-targeted and targeted attack, indicating that our BAST attack improves the transferability of crafted adversarial examples.

\subsection{Analysis of BAST Attack}
\label{sec:strategy}
In this section we study factors which affect the performance of BAST attack. 

We first focus on the ensemble method. Our BAST attack is intuited from bagging and stacking. It is natural to compare it with two special ensemble methods: (1) Without-stacking, which ensembles all models only in the bagging way. Pay attention to the difference of Without-stacking with Bagging attack in Table~\ref{tab:at-least-non-targeted}: Without-stacking is a variant of BAST, also set in at-least-non-targeted protocol, which means the attack on the robust models is always non-targeted. (2) Without-bagging, which ensembles all models only in the stacking way. 

Results are in Table~\ref{tab:compare}. According to the evaluation score, BAST outperforms other two methods in both white-box and black-box protocol. One intriguing phenomenon is that the non-targeted attack success rates of three methods on AdvDeRex are far higher than 15.6\% of non-targeted bagging attack, indicating that performing non-targeted attack on the bagging ensemble weakens the effect of robust models.

We then study the effect of hyperparameters. In one iteration of BAST attack, we conduct targeted attack on easy-to-attack models for $m$ times and then conduct non-targeted attack on robust models for $n$ times. Experiments are conducted on different combinations of $m$ and $n$, and results are shown in Table~\ref{tab:strategy}. We conclude as follows:






(1) Comparing (a) with (b) and (c), 
 the targeted attack success rate of AdvInc-v3 increases with more targeted attack iterations, while the non-targeted attack success rate of AdvDeRex decreases. Thus the trade-off between non-targeted and targeted attack success rate can be adjusted flexibly in this way.
	
(2) Comparing (a) with (d) and (e), on the condition $n=m$, the performance becomes poor as the value of n and m increases. $n$ and $m$ with large value limit the diversity of ensemble. 




\section{Conclusion}

In this paper, we have stated a new adversarial attack mechanism, at-least-non-targeted attack, which is more important for practical application. In order to achieve this, a novel ensemble attack method, BAST attack has been proposed. Extensive experiments have shown the effectiveness of the BAST attack which has improved non-targeted attack success rate while keeping targeted attack performance. That means it outperforms the state-of-the-art ensemble attacks. 

\section{Acknowledgement}

We gratefully thank Tianxiang Ma and Yueming Lv for their assistance with the experiments.

{\small
\bibliographystyle{ieee_fullname}
\bibliography{egbib}
}

\end{document}